\documentclass[conference]{IEEEtran}

\usepackage{chronology}
\usepackage{varwidth}
\usepackage{algorithm}
\usepackage{algpseudocode}
\usepackage{graphicx}
\usepackage{tabularx}
\usepackage{subcaption}
\usepackage{subfig}
\usepackage{tikz}
\usetikzlibrary{spy,arrows,fit,shapes,calc,snakes,trees,decorations.pathmorphing}
\usepackage{amsmath}
\usepackage{multirow}
\usepackage{array}
\usepackage{todonotes}
\usepackage{pgfplots}
\pgfplotsset{compat=1.3}

\pgfdeclarelayer{background}
\pgfsetlayers{main,background}

\begin{document}
\title{TwinCG: Dual Thread Redundancy with Forward Recovery for Conjugate Gradient Methods}

\author{\IEEEauthorblockN{Kiril Dichev}
\IEEEauthorblockA{HPDC\\
School of EEECS\\
Queen's University Belfast\\
Email: K.Dichev@qub.ac.uk}
\and
\IEEEauthorblockN{Dimitrios~S.  Nikolopoulos}
\IEEEauthorblockA{HPDC\\
School of EEECS\\
Queen's University Belfast\\
Email: D.Nikolopoulos@qub.ac.uk}
}

\maketitle

\begin{abstract}
  Even though iterative solvers like the Conjugate Gradients method (CG) have been studied for over fifty years, fault tolerance for such solvers has seen much attention in recent years. For iterative solvers, two major reliable strategies of recovery exist: checkpoint-restart for backward recovery, or some type of redundancy technique for forward recovery. Important redundancy techniques like ABFT techniques for sparse matrix-vector products (SpMxV) have recently been proposed, which increase the resilience of CG methods. These techniques offer limited recovery options, and introduce a tolerable overhead. In this work, we study a more powerful resilience concept, which is redundant multithreading. It offers more generic and stronger recovery guarantees, including any soft faults in CG iterations (among others covering ABFT SpMxV), but also requires more resources. We carefully study this redundancy/efficiency conflict. We propose a fault tolerant CG method, called TwinCG, which introduces minimal wallclock time overhead, and significant advantages in detection and correction strategies. Our method uses Dual Modular Redundancy instead of the more expensive Triple Modular Redundancy; still, it retains the TMR advantages of fault correction. We describe, implement, and benchmark our iterative solver, and compare it in terms of efficiency and fault tolerance capabilities to state-of-the-art techniques. We find that before parallelization, TwinCG introduces around 5-6\% runtime overhead compared to standard CG, and after parallelization efficiently uses BLAS. In the presence of faults, it reliably performs forward recovery for a range of problems, outperforming SpMxV ABFT solutions.
\end{abstract}


\section{Introduction and Related Work}

Iterative solvers like the conjugate gradient method \cite{Hestenes1952}, and its preconditioned variations, are an important and well studied method to solving large systems of linear equations for positive definite matrices.
In the absence of faults, CG shows excellent practical convergence, even if in theory it is susceptible to numerical inaccuracies.

The problem of soft faults, e.g. temporary faults in memory which are not covered by hardware checks, is bound to increase with the number of compute components, due to the decreasing Mean Time Between Failures (MTBF) \cite{Malkowski2010}.

Transient faults are relevant to Conjugate Gradient methods for multiple reasons:
First, due to the very large matrices CG can handle, combined with potentially many iterations, a transient fault may occur.
Second, Conjugate Gradients is in the general case very unstable when such faults occur. 
It is well studied that its convergence can not be guaranteed even for rounding off errors, let alone for transient errors which can affect more significant bits in memory.

In the last five years, there has been a rise in solid research efforts to develop fault-tolerant iterative solvers on the example of CG.
In general, fault-tolerant iterative methods follow one of two important directions, or combinations thereof:

\begin{itemize}
\item Checkpoint-restart
\item Some form of redundancy:
	\begin{itemize}
	\item Time/space redundancy within single-threaded execution (e.g. ABFT sparse matrix-vector product)
	\item Thread redundancy (e.g. Triple Modular Redundancy and Majority Vote) 
	\end{itemize}
\end{itemize}
Unreliable fault tolerance mechanisms do exist (e.g. the self-stabilizing work of \cite{Sao2013}), but we do not focus on them in our approach.
We also remark that we omit a summary of distributed memory versions of CG implementations in this work, since it adds another dimension of complexity; we reserve this direction for future work.

%
%
%
%
%
%

\begin{table}
\centering
\begin{tabular}{|p{0.08\textwidth}|p{0.08\textwidth}|p{0.08\textwidth}|p{0.08\textwidth}|p{0.08\textwidth}|}
\hline
\textbf{Fault-tolerant CG technique} & Online-ABFT \cite{Chen2013} & ABFT SpMxV \cite{Shantharam2012,Fasi2015} 
& TwinCG & Triple Modular Redundancy \\
\hline
\textbf{Redundancy Level} &
None & Redundant data \& computatation & Two redundant threads & Three redundant threads \\
\hline
\textbf{Typical Recovery} & Rollback & Forward (up to 1 SpMxV fault); Rollback integrated in \cite{Fasi2015} & Forward (arbitrary CG faults); Rollback integrated & Forward (arbitrary CG faults) \\
\hline
\end{tabular}
\caption{A comparison of state-of-the-art fault-tolerant CG methods, including TwinCG positioning: Redundancy increases from left to right, and less redundancy generally means less recovery capabilities.}
\label{table:related-work}
\end{table}

We detail and compare the state-of-the-art developments in fault-tolerant CG methods in the following paragraphs; a summary is given in Table \ref{table:related-work}.
We will proceed in increasing level of redundancy, since this is a key aspect of our work as well:

\cite{Chen2013} proposes Online-ABFT, which is a single-threaded CG implementation using checkpoint-restart as fault tolerance approach.
No redundancy of any sort is used, and it is capable of rollback recovery only. To detect faults, Online-ABFT monitors CG properties like residual levels and orthogonality.

The first level of redundancy in work on iterative solvers involves the concept of algorithm-based fault tolerance (ABFT), which was first successfully used by \cite{Huang1984} to detect and correct faults in the matrix-matrix product.
The basic idea is to introduce additional checksums, and some additional computation, which may allow for detection and correction of faults in the matrix-vector product.
The same idea was recently applied to the sparse matrix-vector product in the work of \cite{Shantharam2012} and \cite{Fasi2015}; these contributions improve the resilience of iterative solvers like CG by focusing on the underlying sparse matrix-vector product.
We refer to algorithm-based fault-tolerant versions of sparse matrix-vector product as ABFT SpMxV.
The individual techniques differ in their overhead, and in the capability to recover from faults. The entire mechanism of ABFT SpMxV can be considered an efficient redundancy in a single-threaded execution, which uses some extra space (checksums), and extra time (additional computations).
There are strict limits to what existing ABFT SpMxV can do: the related contributions \cite{Shantharam2012,Fasi2015} detect up to two faults, and correct up to 1 fault.

In addition, the sequential use of redundancy in these approaches, while not significant, always impacts runtime: every additional check is expensive, particularly so for an efficient sparse matrix-vector product.
For example, \cite{Shantharam2012} measures a 7.5 \% overhead for their most efficient approach of detection (but no correction) of faults.

The next step of redundancy is in redundant multithreading, and
this is where our main contribution lies.
Before we outline the few efforts made in this area so far, we list some important advantages of redundant multithreading, compared to non-redundant or less redundant techniques:
\begin{itemize}
\item Compared to non-redundant methods like rollback recovery, every redundancy offers the possibility of forward recovery, always outperforming the former in the presence of faults
\item Compared to less redundant methods like ABFT SpMxV, redundant multithreading is significantly more powerful and generic. It essentially allows to recover from arbitrary types and numbers of faults occurring anywhere in a CG iteration. This easily covers any ABFT SpMxV strategy could do. In fact, existing ABFT SpMxV solution offers limited detection and correction capabilities (up to 1 correction), and it only applies to the sparse matrix-vector product.
\item While redundant multithreading always multiplies the used CPU and cache resources compared to less redundant techniques, it \textit{can} be implemented to only marginally increase the total runtime, since multithreading is inherently suitable for parallelization. We demonstrate this experimentally in this work.
\end{itemize}

A very popular and well-established redundancy technique, which can be implemented in hardware or software, is Triple Modular Redundancy (TMR) (e.g. \cite{Nelson1990}); in this technique, detection of a fault is trivial, and a correction is performed via majority voting of two threads (hopefully) carrying forward correct data.
TMR is the minimal thread redundancy approach recently used for various kernels, including iterative solvers like CG, in the work of \cite{Hukerikar2014,Hukerikar2014b}.
The authors use a holistic compile and runtime system to dynamically spawn redundant threads in certain regions, increasing the number of redundant threads if needed.
The assumption is that the runtime is able to detect certain types of faults (like ECC errors), and dynamically spawn redundant threads for fault tolerance.

The issue with thread redundancy, particularly TMR, is that it uses triple CPU and cache resources that could otherwise be used for more efficient computation, and this is also clearly demonstrated in our experimental results.
As Kanellakis has ably summarized many years before the advent of many-core systems, ``parallel algorithm efficiency implies a minimization of redundancy in the computation, leaving very little room for fault tolerance'' \cite{Kanellakis2013}.

We propose a solution to this problem, by implementing TwinCG, an original fault tolerant CG algorithm, in which we use the minimum possible redundant multithreading, Dual Modular Redundancy (DMR).
DMR is well known, but the novelty is that our solution is capable of forward recovery, similarly to TMR.
We integrate backward and forward recovery, and implement a robust fault tolerance solution in TwinCG.
On one side, we adopt the residual check idea of Online-ABFT for our algorithm.
On the other side, we combine backward/forward recovery, influenced by the Online-Detection/Online-Correction work of \cite{Fasi2015}.

In summary, our contributions to the area of fault-tolerant iterative solvers are:

\begin{itemize}
\item We design and implement an original fault-tolerance algorithm for CG called TwinCG. Its fault tolerance is rooted in using Dual Modular Redundancy
\item TwinCG detects faults, and implements efficient forward recovery from faults, with an intelligent detection and correction process; it can also perform rollback recovery, in the rare cases forward recovery is not deemed possible
\item TwinCG can utilize multi-threaded BLAS libraries like Intel MKL better than more wasteful solutions like Triple Modular Redundancy
\item We implement a more ``relaxed'' version of Online-ABFT, in the sense of reduced sensitivity to insignificant faults; it converges quicker than the original algorithm for the tested problems.
\item We develop a flexible and realistic fault injection mechanism for all implemented solvers, and use it to confirm the fault tolerance of TwinCG for a range of real-world problems
\end{itemize}

The rest of the paper is structured as follows: In section 2 we outline the Conjugate Gradient method, and give a summary of the Online-ABFT, and our modifications. In section 3, we describe in detail our algorithm, TwinCG. In section 4, we analyze the performance of TwinCG, while in Section 5, we evaluate its fault tolerance. We conclude the paper with section 6.

\section{Overview of Conjugate Gradient Methods, and Online-ABFT}

In this section we shortly overview:
\begin{itemize}
\item the CG method, without and with a preconditioner
\item Online-ABFT, since we use some of its principles in our algorithm
\end{itemize}

The Conjugate Gradient method we present here is from \cite{Saad2003}, and refer the reader to the textbook for details.

\begin{algorithm}[H]
\begin{algorithmic}[1]
\State $q_i \gets A * p_i$ \Comment MT SpMxV (Intel MKL)
\State $\alpha_i \gets \frac{\langle r_i, r_i\rangle}{\langle p_i, q_i \rangle} $
\State $x_{i+1} \gets x_i + \alpha_i * {p}_i$
\State $r_{i+1} \gets r_i - \alpha_i * q_i$
\State $\beta_i \gets \frac{\langle r_{i+1} , r_{i+1} \rangle}{\langle r_i, r_i \rangle}$
\State $p_{i+1} \gets r_{i+1} + \beta_i * p_i$
\end{algorithmic}
\caption{Conjugate Gradient}
\label{fig:it-cg}
\end{algorithm}

\begin{algorithm}[H]
\begin{algorithmic}[1]
\State $q_i \gets A * p_i $ \Comment MT SpMxV (Intel MKL)
\State $\alpha_i \gets \frac{\langle r_i, z_i\rangle}{\langle p_i, q_i \rangle}   $
\State $x_{i+1} \gets x_i + \alpha_i * {p}_i$
\State $r_{i+1} \gets r_i - \alpha_i * q_i$
\State $z_{i+1} \gets M^{-1} * r_{i+1}$ \Comment  MT SpMxV (Intel MKL)
\State $\beta_i \gets \frac{\langle r_{i+1} , z_{i+1} \rangle}{\langle r_i, z_i \rangle}$
\State $p_{i+1} \gets z_{i+1} + \beta_i * p_i$
\end{algorithmic}
\caption{Preconditioned Conjugate Gradient}
\label{fig:it-pcg}
\end{algorithm}

The reference implementation of Conjugate Gradients we use is listed in Alg. \ref{fig:it-cg}; the Preconditioned Conjugate Gradients we use is listed in Alg. \ref{fig:it-pcg}.
Our implementation calls the multi-threaded sparse matrix-vector product from Intel MKL, as noted in the pseudocode.
This is central to an efficient implementation of CG since the matrix-vector product dominates the computation time.
We implement PCG using the trivial Jacobi matrix as a preconditioner (that is, $M=diag(A)$ in our prototype).

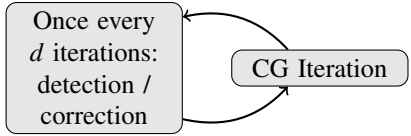
\begin{figure}
  \centering

  \begin{tikzpicture}
\tikzstyle{block} = [rectangle, draw, fill=gray!20, 
    text width=6em, text centered, rounded corners, minimum height=1em, node distance=2.5cm]
\tikzstyle{line} = [draw, -latex']
\node[block](a) {Once every \textit{d} iterations: detection / correction};
\node[block,right of=a,node distance=3cm](b) {CG Iteration};
\path (a) edge [->,thick,bend right] node {} (b)
(b) edge [->,thick,bend right] node {} (a);
  
\end{tikzpicture}
    \caption{Overview of fault-tolerant CG implementations which do correction and detection on CG iteration level: Once every $d$ iterations, faults are detected, and possibly corrected (through backward or forward recovery). At each iteration, textbook CG step is performed.}
    \label{fig:twin-cg}

\end{figure}

The detection and correction steps on the level of CG iterations usually happens once every few iterations, as shown in a simplified version in Fig. \ref{fig:twin-cg}.
This is not the approach taken in ABFT SpMxV solutions, since they only focus on the internals of the matrix-vector product.
However, this is the approach used in our solution.

It is useful to outline the fault-tolerant and non-redundant Online-ABFT algorithm we introduced previously (Table \ref{table:related-work}).
There are multiple reasons we use concepts from this algorithm in our work:
First, it is easy to understand, and introduces no significant changes to CG; this is in strong contrast to the related work on ABFT SpMxV, which requires significant extensions to a matrix-vector product, and we have seen no open-source implementation of any of these.

\begin{figure}
\centering
\begin{tikzpicture}
\tikzstyle{block} = [rectangle, draw, fill=gray!20, 
    text width=6em, text centered, rounded corners, minimum height=1em, node distance=2.5cm]
\tikzstyle{line} = [draw, -latex']

	\node[label=left:{D2}, block, text width=5.5cm, draw=red] (detect-error2) {\begin{varwidth}{5.5cm}Check $ b - A*x_i = r_i $, and $p_{i+1} \perp q_i$ \end{varwidth}};

	\node[label=right:{RR}, block, below right of=detect-error2, text width=2.cm, draw=red] (backward) {\begin{varwidth}{2.cm}Rollback recovery (last checkpoint)\end{varwidth}};
	
	\node[below of=backward,node distance=1.75cm] (end) {};
	\node[below of=detect-error2,node distance=3.5cm] (end2) {};	
	\path[line] (detect-error2) -- node[] {no} (backward);
	\path[line] (backward) --  (end);
	\path[line] (detect-error2) -- node[] {yes} (end2);
\end{tikzpicture}

\caption{Simplified scheme of detection and correction of faults as proposed in Online-ABFT}
\label{fig:chen-recovery}
\end{figure}
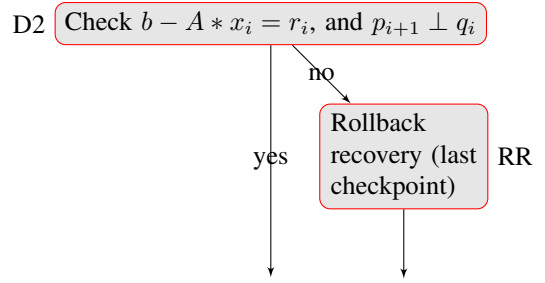

Second, we adopt the residual check of Online-ABFT in our mechanism, as we later explain in detail.
We take the liberty to show a simplified version of Online-ABFT for preconditioned CG methods in Fig. \ref{fig:chen-recovery}.

 The detection step of Online-ABFT consists of two correctness checks:
\begin{itemize}
\item $b - A * x_i = r_i$ holds for each CG iteration
\item The orthogonality $p_{i+1} \perp q_i$ holds for each CG iteration
\end{itemize}

Both of these properties have been studied in earlier work (see e.g. \cite{Saad2003,Shewchuk1994}).
Both of them hold in a fault-free execution, but due to numerical instability these equalities are always approximated (a threshold difference is acceptable).
While we are confident in the soundness of Online-ABFT, we only use the check $b - A * x_i = r_i$ in our reference implementation of Online-ABFT.
The reason for that is that in our experiments, the original Online-ABFT detection, with the threshold levels they use, detected very insignificant faults, which led to disproportionately large amount of checkpoint-restarts, and much slower convergence.
In other words, the original Online-ABFT was too sensitive to faults in our experiments.
Our simplified Online-ABFT version, as can be seen in our experimental results of Table \ref{tab:lambda0.01}, showed good convergence, with less fault sensitivity.

As we mentioned in the introduction, our work is different from Online-ABFT in the use of redundancy, which allows for forward recovery, rather than the rollback recovery in Online-ABFT.
This does not mean that TwinCG will always be a better choice than Online-ABFT, but it is guaranteed to be for high enough fault rates, as will become clear in our evaluation (see Table \ref{tab:lambda0.01}, as well as \cite{Fasi2015}).

\section{TwinCG: Dual Modular Redundancy for CG}
\label{sec:twincg}

\subsection{Overview}
In this section, we propose an original fault-tolerance algorithm for CG, which we call TwinCG.
The algorithm uses dual modular redundancy to detect faults, and unlike other DMR methods is able to recover from transient faults as well.

In most cases, the recovery is efficient forward recovery, as opposed to the more expensive checkpoint-restart recovery.
In the very rare cases where forward recovery is not possible, we still perform a rollback to a checkpointed state.

We call our prototype TwinCG for two reasons:

\begin{itemize}
\item First, our implementation uses two-threaded DMR.
We consider this a minimal extension to single-threaded redundancy, and a less expensive technique than TRM (or any further thread redundancy).
\item Second, both threads, much like twins, perform identical CG iterations (each of them perform the steps shown in Fig. \ref{fig:twin-cg} on replicas of the same data).
Also, they are very supportive of each other: whenever exactly one of them has a severe fault, the healthy thread recovers the faulty one.
\end{itemize}

In the following sections, we carefully describe the detection and correction phases of TwinCG, the logic flow of these phases, and our reasoning behind the design.

\subsection{Detection and Correction}

\begin{figure}
\centering
\begin{tikzpicture}
\tikzstyle{block} = [rectangle, draw, fill=gray!20, 
    text width=6em, text centered, rounded corners, minimum height=1em, node distance=1.25cm]
\tikzstyle{line} = [draw, -latex']

	\node[label={Start thread synchronization}] (begin-synch) {};
	\node[left of=begin-synch, node distance=4.25cm] (begin-synch-left) {};
	\node[right of=begin-synch, node distance=2.5cm](begin-synch-right) {};
	\node[label=left:{D1}, block, below of =begin-synch, text width=4.cm, draw=green] (compare-norm) {\begin{varwidth}{4.cm} $ \left| \left|r^{T1}_i\right| - \left| r^{T2}_i \right| \right|	< \epsilon_1$ \end{varwidth}};

	\node[label=left:{D2}, block, below left of=compare-norm, node distance=2.5cm, text width=3.5cm, draw=red] (detect-error2) {\begin{varwidth}{4.5cm}Check $\frac{\left| b - A*x_i - r_i \right| }{\left| A \right| } < \epsilon_2 $ \end{varwidth}};
		
	\node[block, below right of=compare-norm, node distance=2.5cm, text width=2.5cm] (no-detect-error) {Norms are approx. identical};
	\node[label=left:{FR}, block, below left of=detect-error2, node distance=3.cm, text width=2.5cm, draw=green] (forward) {\begin{varwidth}{2.5cm}Forward recovery (shared memory copy)\end{varwidth}};
	\node[label=left:{RR}, block, below right of=detect-error2, node distance=3.cm, text width=2.cm, draw=red] (backward) {\begin{varwidth}{2.cm}Rollback recovery (last checkpoint)\end{varwidth}};
	\node[below of=forward, node distance=2.5cm] (end-synch) {};
	\node[label=below left:{End thread synchronization}, right of=end-synch, node distance=5.5cm] (end-synch-right) {};
	\node[right of=end-synch-right, node distance=0.15cm] (end-synch-right2) {};
	\node(end-synch-left)[left of=end-synch]{};

	\path[line] (begin-synch) -- (compare-norm);
	\path[line] (compare-norm) -- node[label=left:{\begin{varwidth}{2.cm}'No' for both threads\end{varwidth}}] {} (detect-error2);
	\path[line] (compare-norm) -- node[label=right:{\begin{varwidth}{2.cm}'Yes' for both threads\end{varwidth}}] {} (no-detect-error);
	\path[line] (detect-error2) -- node[label=right:{\begin{varwidth}{2.cm}'No' for both threads\end{varwidth}}] {} (backward);
	\path[line] (detect-error2) -- node[label=left:{\begin{varwidth}{3cm}'No' for exactly 1 thread\end{varwidth}}] {} (forward);
	\path[line] (forward) -- (end-synch);
	\path[line] (no-detect-error) -- (end-synch-right2);
	\path[draw, decorate,decoration={snake,amplitude=.6mm,segment length=2mm,post length=1mm}] (end-synch-left) -- (end-synch-right2);
	\path[draw, decorate,decoration={snake,amplitude=.6mm,segment length=2mm,post length=1mm}] (begin-synch-left) -- (begin-synch-right);
	
	\path[line] (detect-error2) -- (no-detect-error);
\end{tikzpicture}

\caption{Detection and correction logic of each TwinCG thread. The efficient detection step D1, and efficient forward recovery FR, are marked in green. Both are only possible due to redundant multithreading. Their more expensive counterparts are marked in red -- the detection step D2, and the rollback recovery step RR. We apply D2 and RR as our variations on Online-ABFT \cite{Chen2013}}.
\label{fig:recovery}
\end{figure}
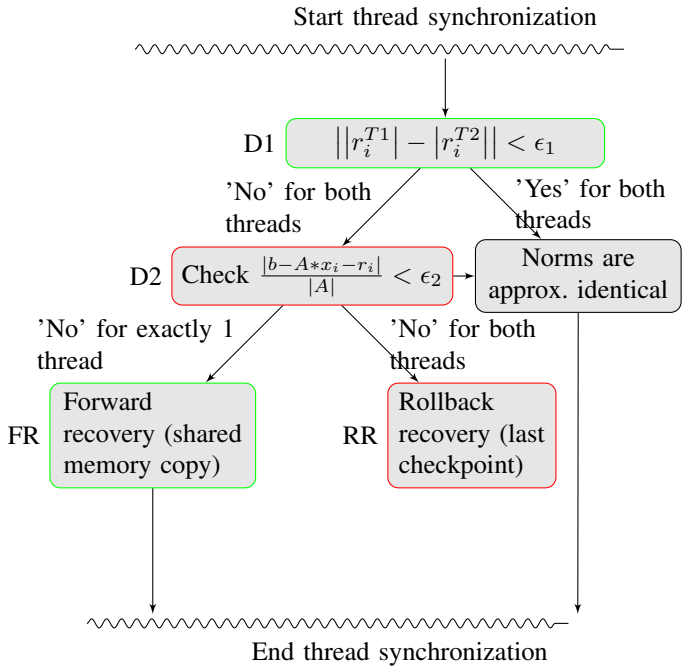

In this part, we detail our implementation of detection and correction of faults for CG, which we display in Fig. \ref{fig:recovery}.

As shown in the diagram, the detection/correction phase, which we perform every \textit{d} iterations, is enclosed by a synchronization window (at the start and at the end of each phase).

The synchronization window is essentially a software-based, and thereof more flexible version of a lock-step execution (see e.g. \cite{Poledna1996}, Ch. 5.4.3).

Our synchronization window is needed for the following reasons:

\begin{itemize}
\item The start synchronization point is needed to ensure threads are in the exact same iteration in order to detect faults correctly
\item The end synchronization point is needed, since the two threads need to read from shared memory (detection steps D1 and D2), and to write into shared memory (correction steps FR and RR); we can not simultaneously allow another thread to perform a write operation as part of continued CG iteration
\end{itemize}

As we see later in this work, the introduction of lock stepping in itself does indeed introduce a few percent overhead.

But in the detection and correction of faults, we make savings in time, especially compared to checkpoint-restart mechanisms.

In the detection/correction phase, we use the redundant computation of each CG iteration at each thread, whenever possible.
As a first detection mechanism (D1), we check the residual norm at each thread:
In the absence of faults, we expect in the ideal case $\left| r^{T1}_i \right| = \left| r^{T2}_i \right|$.
Instead of checking for strict equality, we check that the difference is below a threshold $\epsilon_1$.
This threshold value has important implications, because it becomes a filter for deciding between insignificant and significant bit flips, as visualized in Fig. \ref{fig:ignoring-faults}.
\begin{itemize}
\item If the inequality in D1 holds (always for both threads), either no faults have occurred, or faults have occurred that we consider insignificant. In other words, our use of residual norm in D1 allows us to use $\epsilon_1$ as a significance filter for bit flips which may very often be insignificant \cite{Elliott2013,Gschwandtner2015}. This also has the implication that the two threads may diverge up to a certain point. No correction is needed until then.
\item  If this inequality does not hold (always for both threads), we only know that a significant transient fault has hit at least one, possibly both threads. Since we use dual modular redundancy, we can not use trivial mechanisms like majority vote to decide on where the fault occurs. For this reason, we rely on part of the detection logic of Online-ABFT. We check in parallel at each thread if the condition $r_i = b - A * x_i$ is fulfilled, and denote this detection step as D2. It requires an expensive matrix-vector product, but is only evaluated if detection D1 indicates an issue.
\begin{itemize}
\item If the inequality holds at both threads, we continue without recovery
\item If one thread breaks the inequality, the ''healthy`` thread recovers the ''faulty`` thread via a trivial copy of its CG data in shared memory. This forward recovery is efficient in many ways: it avoids reading a checkpoint (possibly from file), it avoids recomputation, and also importantly, no rolling back to previous iterations of CG is required.
\item If both threads break the inequality, we assume that they both need to be recovered. We resort to a traditional checkpoint-restart, and we roll back a few iterations as Online-ABFT would do. This RR (rollback recovery) step is the most expensive step in the recovery strategy, but it can not be avoided in the improbable case of both threads having significant faults.
\end{itemize}
\end{itemize}

\subsection{Implementation Details}

We implemented TwinCG as a command-line tool in C.
We have an implementation of Conjugate Gradients, and Preconditioned Conjugate Gradients, with the simple Jacobi matrix as a preconditioner.
We use an implementation of File I/O and various operations for sparse matrices in the Compressed Row Storage (CSR) format \cite{CSRformat}.
There are the following external library dependencies:
\begin{itemize}
	\item We use Intel Math Kernel Library for all Sparse BLAS operations
	\item We use GSL \cite{Galassi2009} for the uniform probability distributions, and for the Poisson distribution; we use both for our fault injection mechanism, which we detail later
	\item We use POSIX threads to implement multi-threading, and POSIX condition variables and mutexes to implement the synchronization window
\end{itemize}

The entire implementation thus has no strict requirements on proprietary software.
Intel MKL, as the only proprietary component, can potentially be replaced by an open-source Sparse BLAS implementation.

\subsection{Fault Injection}

We have implemented a new fault injection mechanism on the principle of fully uniform probability distribution, based on the assumption that any bit in memory is equally likely to get flipped.
We believe this assumption to be the most realistic one for transient faults.
It is important to describe the bit flip mechanism in detail, because the fault injection mechanism is an important part of the experimental validation of fault tolerant iterative solvers, and one that is often neglected; it has, however, significant implications to the verification of quality of a fault tolerance scheme.

\begin{figure}
\centering
\begin{tikzpicture}
\tikzstyle{block} = [rectangle, draw, fill=gray!20, 
   text centered, rounded corners, minimum height=1em]
\tikzstyle{line} = [draw, -latex']

\newcommand{\LD}{\langle}
\newcommand{\RD}{\rangle}

\node (0010) [] {$\LD 000\RD$};
\node (0110) [right of=0010,node distance = 2cm] {$\LD 0\textbf{1}0\RD$};
\node (0010b) [below of=0010,node distance=0.5cm] {$\LD 000\RD$};
\node (0110b) [right of=0010b,node distance = 2cm] {$\LD 000\RD$};

\node(filter) at ($(0110)!0.5!(0110b) + (2.5,0)$) {$\left|\LD 0\textbf{1}0 \RD - \LD 000 \RD \right| < \epsilon_1$};
\node (decision) [right of=filter,node distance=3cm] {\begin{varwidth}{1.cm}Ignore bit flip\end{varwidth}};
	\path (0010) edge [->] node[block] {$\lambda$} (0110);
	\path (0010b) edge [->] node[block] {$\lambda$} (0110b);
	\path (0110) edge [->] (filter);
	\path (0110b) edge [->] (filter);
	\path (filter) edge [->] (decision);
	\end{tikzpicture}

\caption{D1 is both a detection mechanism, and filter for insignificant faults: Fault injection with probability $\lambda$ for each thread, and threshold $\epsilon_1$.}
\label{fig:ignoring-faults}
\end{figure}
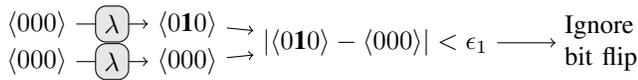

We assume that a bit flip can occur at any bit of any element of the input matrix A.
We do not inject faults in any other element.
It can easily be demonstrated based on a faulty matrix $\widetilde{A}$ in any given iteration that its faults propagate to all CG vectors irreversibly within one iteration;
therefore, we consider this fault injection sufficient (similar to \cite{Sao2013}).
At the end of each iteration, we ``unflip'' any bit flips we may have introduced to the matrix, i.e. we fix the matrix A.
This corresponds to our model of transient faults.

The fault rate $\lambda$ corresponds to how many faults occur per iteration on average, given a Poisson distribution.
Normally, this number should be between 0. and 1.
We then generate a random number of faults per iteration, given a fault rate.
If a fault occurs in an iteration (very rarely more than 1 per iteration), we use uniform distribution to decide which element of the nonzero elements of A gets a bit flip.
We then use again uniform distribution to decide which bit of this element gets flipped.

One consequence of this model is that often the injected faults have no noticeable effect on the convergence of CG.
This effect is well known, and there is relevant work \cite{Elliott2013,Gschwandtner2015} studying how significant is a bit flip in a floating point number, depending on its position in the data representation.
The mantissa of a Binary64 IEEE floating point number holds 52 bits; however, bit flips in the mantissa may often have no significant effect, especially if the exponent is very small.
A fault injection framework using a uniform distribution of bit flips should take this into consideration.
Indeed, in our experimental results, only a fraction of the injected faults can be considered significant (see Fig. \ref{fig:ignoring-faults}).
We still believe this is the right model of fault injection, since it tests fairly and extensively for any bit flips, instead of following subjective decisions serving the purpose of the researcher.

\section{Performance Evaluation of TwinCG}
In this section, we evaluate the performance of TwinCG in following ways:
\begin{itemize}
\item We evaluate the RAM 
\item We carefully describe the thread-to-core pinning we use
\item We compare our two-threaded implementation to a standard CG implementation before memory contention due to SpMxV multithreading
\item We evaluate the effects of enabling SpMxV multithreading in Intel MKL on standard CG, the modified Online-ABFT, TwinCG, and TMR
\end{itemize}

\subsection{Experimental Platform}

For our benchmarks, we use compute nodes on the Kelvin cluster at Queen's. Each node has 2-socket Intel machines with an Intel Xeon  CPU E5-2660 v3 with 2.6 GHz frequency, and 128 GB RAM.
Each of the two sockets has 10 cores, which share per socket 25 MB L3 cache, with 256 KB L2 cache, and 32 KB L1 cache per core.
Intel Hyper-Threading was disabled on the nodes, so there are 20 physical and virtual cores available per node.

\subsection{Main Memory Footprint}

For its backward/forward recovery strategy, TwinCG requires:
\begin{itemize}
\item triple storage for all CG vectors (see Alg. \ref{fig:it-cg} and \ref{fig:it-pcg}). Each thread holds a copy of the vectors for forward recovery, and they have shared access to another copy of all CG vectors as a checkpoint for backward recovery
\item duplicate storage for the problem matrix (each thread holds a copy for forward recovery); a third copy of the matrix is not required as a checkpoint for backward recovery, because we assume only transient (that is, temporary) faults can affect the matrix; these faults only permanently affect the CG vectors, not the matrix.
\end{itemize}
 
We use the efficient Compressed Row Storage format commonly used for sparse matrices.
The amount of storage needed is proportional to the number of nonzeroes (nnz).
We profiled the heap use for various problems.
We experimented with the range of matrices we use for most experiments, as listed in Table \ref{tab:problems}, where the nnz varies between 340 thousand and 7.6 million.
For our largest test matrix (G3\_circuit with 7.6 mio nnz), our implementation consistently used 488.5 MB of heap memory with standard CG, and 727 MB of heap memory for the two-threaded TwinCG.
This includes the entire allocated memory for all data structures.
We therefore consider the problem of allocating a few extra hundred MB of additional RAM memory acceptable.

\subsection{Efficient Thread-to-Core Mapping in TwinCG}

\begin{figure}
\centering
\begin{tikzpicture}
\tikzstyle{block} = [rectangle, draw, fill=gray!20, 
    text width=6em, text centered, rounded corners, minimum height=1em, node distance=2.cm]
\node[block](main){TwinCG main};
\node[block,above right of=main] (t1) {POSIX thread for CG};
\node[block,right of=t1,text width=3em] (mkl1) {Intel MKL};
\node[block,right of=mkl1] (omp1) {OpenMP threads};
\node[block,below right of=main] (t2) {POSIX thread for CG};
\node[block,right of=t2,text width=3em] (mkl2) {Intel MKL};
\node[block,right of=mkl2] (omp2) {OpenMP threads};

\node[below of=omp1,node distance=0.8cm] (sock1-label) {Socket 1};
\node[below of=omp2,node distance=0.8cm] (sock2-label) {Socket 2};
%
\path[draw] (main) -- (t1) -- (mkl1) -- (omp1);
\path[draw] (main) -- (t2) -- (mkl2) -- (omp2);

\begin{pgfonlayer}{background}
\path[fill=gray!60,rounded corners, draw=black!50, dashed, opacity=0.2]
      ($(omp1) + (-1.25,-1)$) rectangle ($(omp1) + (1.5,1)$);
\path[fill=gray!60,rounded corners, draw=black!50, dashed, opacity=0.2]
      ($(omp2) + (-1.25,-1)$) rectangle ($(omp2) + (1.5,1)$);      
\end{pgfonlayer}

\end{tikzpicture}
\caption{Illustration of nested thread paralellism in TwinCG, with each redundant POSIX thread calling the OpenMP-parallel SpMxV routine in Intel MKL. We also outline the thread-to-socket pinning we experimentally find to be most efficient.}
\label{fig:nested-threading}
\end{figure}
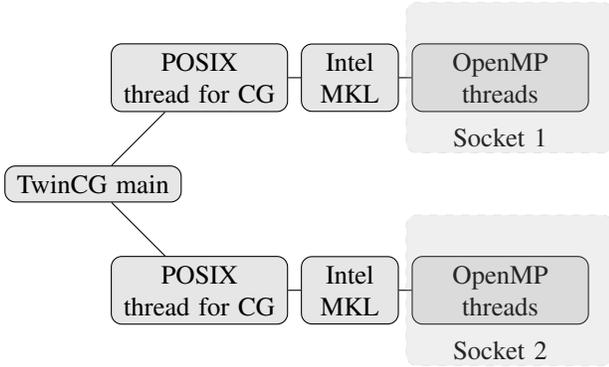

The thread-to-core mapping in TwinCG is very important, and a bad mapping has a detrimental effect on performance, especially for a memory-bound problem like CG.

In our implementation, there is a two level hierarchy of threads, as shown in Fig. \ref{fig:nested-threading}:
\begin{itemize}
  \item The main program first spawns POSIX threads for CG (1 for standard CG or Online-ABFT, 2 for TwinCG, or 3 for Triple Modular Redundancy).
  \item Each CG kernel calls SpMxV, which is implemented in Intel MKL, and is internally parallelised using OpenMP threads.
\end{itemize}

While Intel MKL is thread-safe for multithraded applications to use, the thread-to-core mapping when using POSIX threads to call OpenMP threads is not trivial.
The challenges and solutions are described in detail in \cite{IntelMKLandPthreads}.
We are unable to properly set the thread affinity without source code modifications.
The reason is that the OpenMP runtime only has a global affinity view, but no notion of the POSIX thread which runs an OpenMP-parallel region.
However, pinning down threads depending both on the first-level POSIX thread number, and second-level OpenMP thread number, is possible with some code modifications.
The affinity can be set e.g. via Intel or POSIX Thread Affinity API.
We use the Intel API, specifically \verb1kmp_set_affinity1 calls.

The thread-to-core mapping for TwinCG we choose fits very well with the used 2-socket platform: We place each POSIX thread, and all its associated OpenMP threads, on its own dedicated socket (the OpenMP thread groups are thus pinned to two different sockets, as shown in Fig. \ref{fig:nested-threading}).
The motivation is to avoid memory contention between the two redundant threads, which mostly use independent data sets.

For the TMR version, we use the same strategy for the first two threads, and pin the OpenMP threads of the third POSIX thread consecutively to sockets one and two.
An ideal solution for TMR in this setting does not exist:  we have two sockets available, and need to utilize three redundant threads, each of them using a number of OpenMP threads computing the memory-intense SpMxV.
We can not avoid memory contention in this scenario.

\subsection{CPU Footprint}

\subsubsection{Evaluation of Multithreading and Synchronization in TwinCG}

It is important to evaluate the effects of multithreading in our two-threaded implementation of CG.
Two important questions arise: 
\begin{itemize} 
\item Is there a scenario where no memory contention occurs, and the efficiency of TwinCG in wallclock time is not affected compared to single-threaded CG?
\item For this scenario, how significant is the overhead of the synchronization window we presented?
\end{itemize}

To begin with, we experimentally confirm that memory contention can be avoided with following setting: use of a single-threaded Intel MKL SpMxV, and pinning of the two redundant POSIX threads, and the derived OpenMP threads, on different sockets to avoid competition for L3 cache.

After performing this setup, we run benchmarks with a number of matrix sizes, ranging from 340 thousand nonzeros to 7.7 mio. nonzeros.
No faults are injected for both cases.
Each run only terminates after convergence ($tol = 10^{-10}$).
The average time over 10 iterations of each experiment was used for standard CG.
For TwinCG, the maximum time of the two threads was taken for each run, and this maximum was averaged over 10 runs.

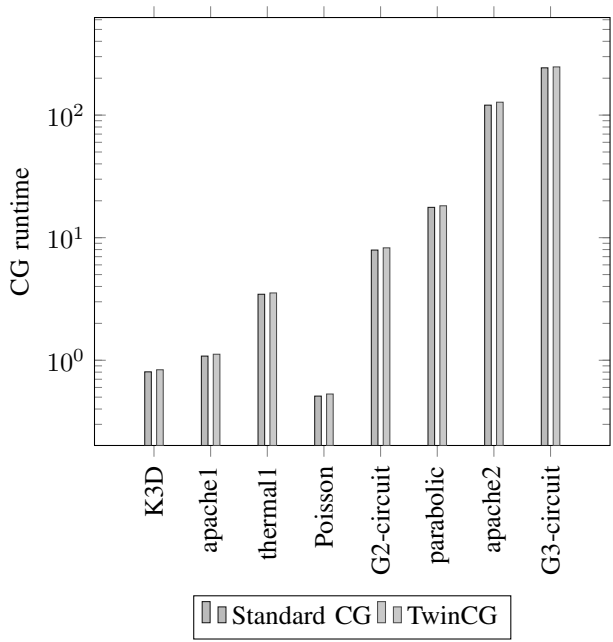
\begin{figure}
\centering
\begin{tikzpicture}[]
\selectcolormodel{gray}
\begin{axis}[ymode=log,x tick label style={rotate=90},ybar,enlargelimits=0.15,legend style={at={(0.5,-0.35)},anchor=north,legend columns=-1},ylabel={CG runtime},symbolic x coords={K3D,apache1,thermal1,Poisson,G2-circuit,parabolic,apache2,G3-circuit},xtick=data,nodes near coords align={vertical},bar width=2.5]
\addplot coordinates {(K3D,0.804386) (apache1,1.08117) (thermal1,3.45463) (Poisson,0.509913) (G2-circuit,7.92556) (parabolic,17.6754) (apache2,120.565) (G3-circuit,243.044)};
\addplot coordinates {(K3D,0.836466) (apache1,1.1195) (thermal1,3.54124) (Poisson,0.53031) (G2-circuit,8.26252) (parabolic,18.2096) (apache2,127.413) (G3-circuit,247.279)};
\legend{Standard CG,TwinCG}
\end{axis}

\end{tikzpicture}
\caption{Comparison of standard CG and TwinCG using sequential Intel MKL. y-axis measures (in log scale) the average CG time to convergence for standard CG and for TwinCG (two-threaded). When using proper thread-to-core pinning, we see at most 5-6\% overhead for TwinCG, which we attribute to lock stepping (Fig. \ref{fig:recovery}) across the two redundant threads.}
\label{fig:overhead-twincg}
\end{figure}

Fig. \ref{fig:overhead-twincg} shows the time to convergence for standard CG, and for TwinCG.
Standard CG performs only CG iteration time, while TwinCG also includes the lock stepping time.
When using proper thread-to-core pinning as described previously, the actual CG computation time, excluding the lock stepping overhead of TwinCG, is comparable across both single-threaded and two-threaded versions. 
TwinCG did not introduce a computation overhead, even for the largest problems.
However, the two TwinCG threads did slightly differ in their compute time across iterations, and this resulted in the lock stepping phase (once every 5 iterations) forcing the quicker thread to wait.
We profiled this TwinCG overhead in total, and found that it amounts to at most 5.5\%  compared to the CG computation phase.
This overhead directly determined the total TwinCG overhead compared to standard CG.
It is unclear if this TwinCG overhead can be eliminated; it may be due to inefficiencies in our lock stepping implementation, or a CPU/cache utilization difference for redundant computation on different sockets.

\subsubsection{Combining Redundant Multithreading and BLAS Multithreading}

Redundant multithreading for TwinCG, and an efficient multithreaded SpMxV kernel as we use it, poses a conflict for resources for orthogonal purposes: fault tolerance, and efficiency.
This conflict is intuitively clear, but its experimental validation is tricky, and requires precise pinning of threads to cores to avoid the significant effects of cache sharing for all tested settings.
We have described our optimal choice of thread affinity, and we present experimental results based on this choice.

\begin{figure}
\includegraphics[width=0.5\textwidth]{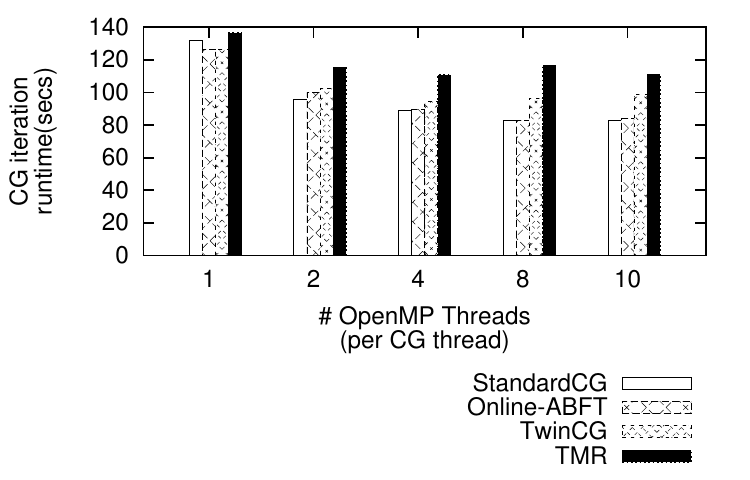}
\includegraphics[width=0.5\textwidth]{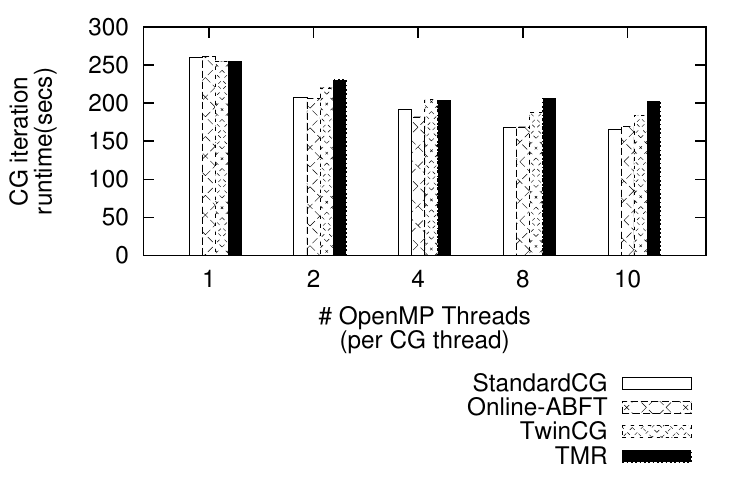}
\caption{Effect of using both redundant threads for fault tolerance, and multi-threaded SpMxV operations via MKL threads. Problems shown are apache2 (above) and G3\_circuit (below). Standard CG and Online-ABFT do not use any redundancy, and therefore can better explore SpMxV multithreading, especially for larger MKL counts. TwinCG can still explore multithreaded SpMxV with very little loss in performance (with efficient thread-to-core pinning); TMR experiences memory bottlenecks for larger OpenMP thread count (per CG thread).}
\label{fig:combining-threads}
\end{figure}

We show experimental results in Fig. \ref{fig:combining-threads}, which shows the time to solution, including only the accumulated CG iteration time for simplicity.
We add here that our detection/correction times consistently outperformed those for Online-ABFT, since the lock stepping of our solution is compensated for by detection step D1, which is more efficient than D2.
The point of this plot, however, is to show the conflict between redundancy and BLAS efficiency.
We have compared our solution TwinCG with our reference implementation of standard CG, Online-ABFT, and Triple Modular Redundancy.
The shown problems are the largest two, apache2 (around 5 mill nnz), and G3\_circuit (7.6 mill nnz); we choose larger problems for more representative results.
The x axis represents the OpenMP threads we set \textit{per CG thread} for any implementation, whether it is single-threaded or multi-threaded.
The y axis represents the total CG computation time, excluding detection; while our detection is more efficient than e.g. Online-ABFT, we leave it out of the plot for simplicity.

We expect standard CG to exploit parallelism best, since no redundancy is used for any fault tolerance.
We also expect Online-ABFT to exploit parallelism well, since its redundancy is sequential, and does not inhibit the BLAS parallelism.
Both of these expectations are confirmed: StandardCG and Online-ABFT perform well overall with increased BLAS parallelism.
We then expect our two-threaded redundancy to outperform the TMR solution clearly for large OpenMP threads; this is due to both L3 cache contention, and ultimately oversubscription (for 8 and more OpenMP threads per CG thread) for TMR.
This expectation is also confirmed.

In summary, these results fully mirror the increasing level of redundancy as summarized in Table \ref{table:related-work}.
TwinCG, while offering a much broader and generalized recovery spectrum than Online-ABFT, or any ABFT SpMxV approach, comes very close in performance to it.
However, Online-ABFT does have an advantage for high BLAS parallelism, since the thread redundancy of TwinCG limits the available cores for thread parallelism on the BLAS level.

\subsection{On Simultaneous Multithreading}

Another related topic, which focuses on the architecture aspect of multithreading, is the Simultaneous Multithreading Architecture \cite{Tullsen1995} (of which Intel Hyper-Threading is an example).
We do not see promising results when employing hyper-threading for redundant threads on the same physical core; this is not surprising, considering that even pinning two redundant threads on the same physical socket has detrimental effects for performance (see TMR in Fig. \ref{fig:combining-threads}).
In general, hyper-threading is not recommended with Intel MKL in \cite{intel-mkl-and-hyperthreading}, so also exploring hyper-threading for each SpMxV call is unlikely to yield any benefits.

\section{Fault Tolerance of TwinCG}
We previously evaluated the performance of TwinCG without faults, and without verifying in any way that the proposed scheme offers fault tolerance capabilities.
In this section, we verify using a range of problems that in theory and in practice, TwinCG should be able to recover from the majority of faults using the efficient forward recovery familiar from mechanisms like TMR.

\subsection{Expected Gains}

In this section we somewhat formalize why TwinCG is efficient in both its detection, and in its recovery.
First, its detection rarely resorts to the expensive D2 step.
Second, it can recover from significant faults almost exclusively via forward recovery, and rarely needs backward recovery.
This result is further confirmed in our experimental validation.

We base our formalization on following setting:
\begin{itemize}
\item the fault rate $0 < \lambda < 1$ as a mean fault rate per thread, and per CG iteration (reciprocal to the Mean Time Between Failure).
\item we perform a detection step once every 5 iterations, same as Online-ABFT.
\end{itemize}

\subsubsection{Fault rate}
Redundant multithreading increases the likelihood of a fault compared to standard CG, or Online-ABFT, since each thread operates on its own data, which may experience bit flips.
The likelihood of a fault in an iteration of Online-ABFT is exactly $\lambda$.
The likelihood of a fault in our TwinCG method is $ 1-(1-\lambda)^2 $.
E.g., for $\lambda = 0.01 $, Online-ABFT experiences a fault in an iteration with 1\% probability, while TwinCG experiences a fault with 1.99\% probability.

\subsubsection{Detection}

The detection step D1 of Fig. \ref{fig:recovery} comes with no overhead other than the synchronized window required for the entire phase. Both threads already hold a copy of $\left| r_i \right|$.
D1 holds if we have no fault in any of 2 threads within 5 iterations; the probability of that is $(e^{-\lambda})^{10}$. 
E.g., if $\lambda = 0.01 $, with probability of ca. 90\% we have no faults within 5 iterations, and avoid the more expensive detection D2 altogether.
In comparison, we would perform the expensive detection step D2 every 5 steps anyway in Online-ABFT, or the derived ABFT SpMxV methods of \cite{Fasi2015}.

\subsubsection{Correction}
\label{sec:estimated-correction}
Let us assume we have detected that at least one fault has occurred, i.e. step D1 has returned negative result, and we need to perform detection step D2 at each thread (in parallel).

The probability of exactly 1 of 2 threads experiencing at least 1 fault within 5 iterations is $2 * e^{(-5*\lambda)} * (1-e^{(-5*\lambda)})$, and based on following factors:
\begin{itemize}
\item with probability $e^{(-5*\lambda)} $ thread 1 experiences no faults in 5 iterations
\item with probability $(1-e^{-5*\lambda})$ thread 2 expriences at least 1 fault in 5 iterations.
\item the factor two is for the reversed scenario.
\end{itemize}

From the same logic, the probability of both threads experiencing a fault with d=5 (detection each 5 iterations) is $(1 - e^{(-5*\lambda)}) ^ 2$.

Based on this derivation, it follows that e.g. for a fault rate of $\lambda = 0.01$, the probability for forward recovery for d=5 (detection each 5 iterations) is around 9.2\%.
For the same fault rate, the probability for backward recovery is around 0.2\%.

This comes with all the advantages of forward recovery; the most significant advantage is that not a single CG iteration is wasted.
The healthy thread restores the faulty thread, and both threads resume computation in the next iteration step.
A rollback is by far the costliest step in approaches like Online-ABFT.

\subsection{Experimental Setup}

In this section, we will evaluate the fault tolerance of TwinCG in regard to transient faults.
In the absence of transient faults, we converge in the same number of iterations as standard CG.
As we have outlined in Sect. \ref{sec:twincg}, we expect to be able to perform efficient forward recovery without losing iteration steps if at most 1 thread is hit by a significant fault;
if both threads are hit by a significant fault, we perform backward recovery.

The parameters we use are:
\begin{itemize}
\item a detection is triggered each 5 iterations, and a backup is performed each 10 iterations (as in Online-ABFT)
\item $e_1 = 10^{-15}$, $e_2 = 10^{-10}$ (See Fig. \ref{fig:recovery})
\item We use $e = 10^{-10}$ for our version of Online-ABFT, for compatibility with the original, and in agreement with $e_2$ of TwinCG
\item We use $tol = 10^{-10}$ for TwinCG
\item We abort each run if no convergence is reached at 6000 iterations, and mark this down (rightmost column in Table \ref{tab:lambda0.01}).
\end{itemize}

The problems we use are all from the freely available University of Florida Sparse Matrix Collection \cite{Davis2011}, with the exception of the better conditioned K3D problem \cite{Sao2013}.
We list the problems, their nonzero count, and their condition number estimate (according to MATLAB) in Table \ref{tab:problems}.
Our main criteria was to have a good range of real-world problems, which converge within 6000 iterations, in a fault-free execution.

\begin{table}
\centering
\begin{tabular}{|c|c|c|}
\hline
Problem & nnz & condition number \\
\hline
K3D & 340200 & $645$ \\
apache1 & 542184 & $4 * 10^6$ \\
thermal1 & 574458 & $5 * 10^5$ \\
Pres\_Poisson & 715804 & $3.2 * 10^6$ \\
G2\_circuit & 726674 & $2 * 10^7$ \\
parabolic\_fem & 3674625 & $2.1 * 10^5$ \\
apache2 & 4817870 & $ 5.3 * 10^6 $ \\
G3\_circuit & 7660826 & $ 2.24 * 10^7 $ \\
\hline
\end{tabular}
\caption{Selection of problems used in our benchmarks and fault tolerance tests.}
\label{tab:problems}
\end{table}

We can experiment with arbitrary fault injection rates $\lambda$. 
We here show results for $\lambda = 0.01 $ and $ \lambda = 0.1 $, which corresponds to an average of 1 random bit flip per thread every 100 iterations, or 1 random bit flip per thread every 10 iterations.
As outlined earlier, many bit flips need not lead to faults, depending on where they occur; similarly, some problems may not be ill-conditioned enough to be easily affected by bit flips.
We tested all problems presented in Table \ref{tab:problems}.
Apart from our implementation TwinCG, we experiment with reference implementations of standard PCG (no fault tolerance), and Online-ABFT (fault tolerance with no redundancy).

\begin{table}
\centering
\begin{tabularx}{0.5\textwidth} {|c|c|c|c|c|c|}

  \hline
 \begin{varwidth}{0.75cm}Problem\end{varwidth} 
 & 
 \begin{varwidth}{2cm}Iterative solver\end{varwidth}
 &
 \begin{varwidth}{1cm}\# RR\end{varwidth}
 &
 \begin{varwidth}{1cm}\# FR\end{varwidth}
 &
 \begin{varwidth}{1cm}\# Iter\end{varwidth} &
  \begin{varwidth}{0.25cm}\% abor-ted (6K iter)\end{varwidth}
 \\
\hline
\multicolumn{6}{|c|}{$\lambda = 0.01 $}
\\
\hline
\multirow{3}{*}{K3D}
& StandardCG & - & - & 2782.62 & 28\%
\\
& Online-ABFT & 0.6 & - & 1067.63 & 0\%
\\
& TwinCG & 0 & 14.35 & 1045.13 & 0\%
\\
\hline
\multirow{3}{*}{apache1}
& StandardCG & - & - & 548.5 & 0\%
\\
& Online-ABFT & 0 & - & 548.12 & 0\%
\\
& TwinCG & 0 & 0.17 & 548.13 & 0\%
\\
\hline
\multirow{3}{*}{thermal1}
& StandardCG & - & - & 1731.82 & 0\%
\\
& Online-ABFT & 0.4 & - & 1520.17 & 0\%
\\
& TwinCG & 0 & 12.4 & 1516.97 & 0\%
\\
\hline
\multirow{3}{*}{Pres\_Poisson}
& StandardCG & - & - & 1298.18 & 10\%
\\
& Online-ABFT & 0.3 & - & 754.85 & 0\%
\\
& TwinCG & 0 & 6.25 & 752.283 & 0\%
\\
\hline
\multirow{3}{*}{G2\_circuit}
& StandardCG & - & - & 2502.98 & 0\%
\\
& Online-ABFT & 0.4 & - & 2365.68 & 0\%
\\
& TwinCG & 0 & 2.57 & 2324 & 0\%
\\
\hline
\multirow{3}{*}{parabolic\_fem}
& StandardCG & - & - & 1519.9 & 3\%
\\
& Online-ABFT & 0.32 & - & 1063.33 & 0\%
\\
& TwinCG & 0 & 0.6 & 1061 & 0\%
\\
\hline
\multirow{3}{*}{apache2}
& StandardCG & - & - & 5809.08 & 63\%
\\
& Online-ABFT & 0.87 & - & 5498.82 & 3\%
\\
& TwinCG & 0 & 9.95 & 5456.02 & 0\%
\\
\hline
\multirow{3}{*}{G3\_circuit}
& StandardCG & - & - & 4503.1 & 0\%
\\
& Online-ABFT & 0.43 & - & 4499.33 & 0\%
\\
& TwinCG & 0 & 2.88 & 4496 & 0\%
\\
\hline
\multicolumn{6}{|c|}{$\lambda = 0.1 $}
\\
\hline
\multirow{3}{*}{K3D}
& StandardCG & - & - & 5925.45 & 98\%
\\
& Online-ABFT & 10.27 & - & 1269.35 & 0\%
\\
& TwinCG & 0.2 & 70.73 & 1112.62 & 0\%
\\
\hline
\multirow{3}{*}{apache1}
& StandardCG & - & - & 738.55 & 1\%
\\
& Online-ABFT & 0.05 & - & 619.97 & 0\%
\\
& TwinCG & 0 & 1.47 & 549.18 & 0\%
\\
\hline
\multirow{3}{*}{thermal1}
& StandardCG & - & - & 3276.57 & 13\%
\\
& Online-ABFT & 4.65 & - & 1552.93 & 0\%
\\
& TwinCG & 0.08 & 38.27 & 1517.83 & 0\%
\\
\hline
\multirow{3}{*}{Pres\_Poisson}
& StandardCG & - & - & 5568.6 & 91\%
\\
& Online-ABFT & 1.85 & - & 767.02 & 0\%
\\
& TwinCG & 0 & 20.78 & 752.65 & 0\%
\\
\hline
\multirow{3}{*}{G2\_circuit}
& StandardCG & - & - & 3135.65 & 0\%
\\
& Online-ABFT & 4.25 & - & 2580.2 & 0\%
\\
& TwinCG & 0.35 & 21.58 & 2336.6 & 0\%
\\
\hline
\multirow{3}{*}{parabolic\_fem}
& StandardCG & - & - & 5127.77 & 81\%
\\
& Online-ABFT & 2.9 & - & 1083.08 & 0\%
\\
& TwinCG & 0.03 & 5.65 & 1061.25 & 0\%
\\
\hline
\multirow{3}{*}{apache2}
& StandardCG & - & - & 6000 & 100\%
\\
& Online-ABFT & 9.27 & - & 5876.17 & 60\%
\\
& TwinCG & 0.25 & 61.48 & 5474.27 & 3\%
\\
\hline
\multirow{3}{*}{G3\_circuit}
& StandardCG & - & - & 4550.67 & 0\%
\\
& Online-ABFT & 5.43 & - & 4536.73 & 0\%
\\
& TwinCG & 0.08 & 27.133 & 4496.83 & 0\%
\\

\end{tabularx}

\caption{Validation of fault tolerance of TwinCG, using standard CG and the simplified Online-ABFT version for comparison. We give results for $ \lambda = 0.01 $ and $\lambda = 0.1 $. Every field value is averaged over 60 iterations. We choose to terminate CG iterations at maximum 6000, in which case we abort a run (rightmost column shows \%).}
\label{tab:lambda0.01}
\end{table}

The compiled results are shown in Table \ref{tab:lambda0.01}.
We list the behaviour of standard CG, our version of Online-ABFT, and TwinCG, in terms of iterations to converge (if convergence is reached in 6000 iterations), and the performed backward and/or forward recovery.
Every single number is an average over 60 independent executions, each triggering different bit flip patterns for the shown $\lambda$ probabilities.
Rightmost column shows \% of aborted runs after 6000 iterations.
Standard CG makes no recovery efforts; Online-ABFT implements checkpoint-restart, i.e. backward recovery; TwinCG mainly performs forward recovery, and is also able to perform backward recovery.

Looking at Table \ref{tab:lambda0.01}, we first observe that the problems K3D and apache2 are most sensitive to bit flips, followed by Pres\_Poisson and parabolic\_fem.
Standard CG rarely converges for $\lambda = 0.1 $ for either of these 4 problems.
For all tested problems, TwinCG realiably recovers forward, and converges in less iterations than Online-ABFT.
This proportionally translates in faster execution time.
We expect this overall faster convergence for TwinCG, since it can use forward recovery, and this is experimentally verified here.
Note that each time rollback recovery is needed (\# RR), this corresponds to either 5 or 10 iterations of rollback.
The displayed results make it unnecessary to draw a comparison to the more expensive TMR method, which in its forward recovery would behave similarly to TwinCG.
The rollback recovery (via checkpoint) in TwinCG is very rare (for double faults), as was estimated in Sect. \ref{sec:estimated-correction}.

\section{Conclusion}
In this work, we introduced a fault-tolerant implementation of Conjugate Gradient methods, TwinCG, which uses two redundant threads, and is able to not just detect, but also correct, transient faults.
The recovery is powerful, in the sense that it covers arbitrary faults occurring either in the most expensive SpMxV operation, or anywhere else within a CG iteration, and it also can easily be extended to any soft faults.
Of course, in terms of CPU usage, our redundant multithreading doubles the used CPU and cache resources compared to a single-threaded CG implementation, since the CG calculation is duplicated, but does not triple them as TMR.
We further evaluated the performance of our implementation in detail, and concluded that the introduced lock stepping across two threads adds up to 6\% overhead relative to the baseline, but no other overhead is introduced compared to a single-threaded CG implementation, in terms of wallclock time.
To achieve this performance, we use a proper thread-to-core pinning.
TwinCG explores BLAS parallelism very well, but its baseline is slightly less efficient than this of less robust and non-redundant solutions like Online-ABFT, which can explore more available cores.
We then tested the fault tolerance of TwinCG on a number of problems, and concluded that its ability to recover from faults shows all the strengths that TMR solutions have, almost exclusively performing forward recovery and outperforming therefore algorithms like Online-ABFT.

In summary, we see TwinCG as a more efficient, and equally robust, alternative to Triple Modular Redundancy solutions for Conjugate Gradients methods, with the same powerful recovery capabilities, extending recovery from transient faults beyond ABFT SpMxV solutions.
TwinCG is particularly efficient for settings where different soft faults and higher fault rates in CG iterations may be anticipated.

\section*{Acknowledgment}
This project has received funding from the European Union’s Horizon 2020 research and innovation programme under grant agreement No 671603.

We would like to thank Enrique Quintana Ort\'i and Jos\'e I. Aliaga for providing an initial implementation of CG with Intel MKL and CSR format support, and for answering all related questions.
We are also thankful to Giorgis Georgakoudis and Charalampos Chalios for the many insights into cache hierarchies and OpenMP parallelism they provided.

\bibliographystyle{IEEEtran}
\bibliography{references}
\end{document}